\documentclass[preprint,showpacs,preprintnumbers,amsmath,amssymb]{revtex4-1}
\usepackage{graphics,epsfig,subfigure}
\usepackage{diagbox}
\usepackage[usenames]{color}
\usepackage[colorlinks,
            linkcolor=blue,
            anchorcolor=red,
            citecolor=red
            ]{hyperref}
\usepackage{color}
\usepackage{graphicx}
\usepackage{amsmath}
\usepackage{amsfonts}
\usepackage{amssymb}
\usepackage{txfonts}
\usepackage{indentfirst}
\usepackage{booktabs}

\begin{document}
\renewcommand{\baselinestretch}{1.3}
\newcommand\beq{\begin{equation}}
\newcommand\eeq{\end{equation}}
\newcommand\beqn{\begin{eqnarray}}
\newcommand\eeqn{\end{eqnarray}}
\newcommand\nn{\nonumber}
\newcommand\fc{\frac}
\newcommand\lt{\left}
\newcommand\rt{\right}
\newcommand\pt{\partial}

\title{Rotating multistate boson stars}
\author{  Hong-Bo Li\footnote{lihb2017@lzu.edu.cn}, Shuo Sun\footnote{sunsh17@lzu.edu.cn}, Tong-Tong Hu\footnote{hutt17@lzu.edu.cn}, Yan Song\footnote{songy18@lzu.edu.cn} and Yong-Qiang Wang\footnote{yqwang@lzu.edu.cn, corresponding author
}
}

\affiliation{Research Center of Gravitation and  Institute of Theoretical Physics  and  Key Laboratory for Magnetism and Magnetic of the Ministry of Education, Lanzhou University, Lanzhou 730000, China}

\begin{abstract}
In this paper, we  construct  rotating  boson stars  composed of the coexisting states of two scalar fields, including the ground and first excited states. We show the coexisting phase with both the ground and  first excited states for rotating multistate boson stars. In contrast to the solutions of the nodeless boson stars, the rotating boson stars with two states have two types of nodes, including the $^1S^2S$ state and the $^1S^2P$ state.
Moreover, we explore the properties of the  mass  $M$ of rotating boson stars with two states as a function of the synchronized frequency $\omega$, as well as  the nonsynchronized frequency $\omega_2$.  Finally, we also study the dependence of the mass $M$ of rotating boson stars with two states on angular momentum for both the synchronized frequency $\omega$ and  the nonsynchronized frequency $\omega_2$.
\end{abstract}

\maketitle

\section{Introduction}\label{Sec1}

In the mid-1950s, John Wheeler found the classical fields of electromagnetism coupled to the Einstein gravity theory \cite{Wheeler:1955zz,Power:1957zz}.
In the next half century,  Kaup et al.  \cite{PhysRev.187.1767} replaced electromagnetism with a free, complex scalar field and found Klein--Gordon geons \cite{Kaup:1968zz} that have become well-known as boson stars (BSs).

Firstly, boson stars that were constructed with fourth and sixth power $|\phi|$-term potentials were considered in  \cite{Mielke:1981}, and there is a  more detailed analysis of a potential with only the quartic term in Ref.  \cite{1986PhRvL..57.2485C}. Moreover, by using a V-shaped potential proportional to   $|\phi|$, one can also find the compact boson stars  \cite{Hartmann:2012da}, and the same V-shaped potential with an additional quadratic massive term has also been studied in  \cite{Kumar:2015sia}. In Ref.  \cite{Guenther R.L:1995}, the Newtonian boson stars were investigated, and  the boson field coupled to an electromagnetic field in Ref. \cite{Jetzer P:1989}. Furthermore, the study of boson stars can be extended to the boson nebulae charge  \cite{Dariescu C:2010,Murariu G.:2010,Murariu G.:2008}, the charged boson stars with a cosmological constant case \cite{Kumar:2016sxx}, and the charged, spinning Q-balls case \cite{Brihaye:2009dx}. In addition, the fermion-boson stars were studied in Refs.  \cite{Henriques:1989,Henriques:1990,deSousa:1995ye,Pisano:1995yk}.
Most of the studies of the solutions have focused on the model of one scalar hair with the fundamental solutions. Recently, the spherically symmetric, nonrotating  boson stars with two coexisting states  was discussed in Refs. \cite{Bernal:2009zy,UrenaLopez:2010ur}, which combined the ground state  with the first excited state, and the study of the case of nonrotating boson stars with two coexisting states can be extended numerically to the phase shift and  dynamics \cite{Hawley:2002zn},
which is the individual particlelike configurations for each complex field case \cite{Brihaye:2009yr,Brihaye:2008cg}. Besides, the axisymmetric rotating radially excited boson stars has been studied  in \cite{Collodel:2017biu} and see Ref. \cite{Liebling:2012fv} for a review.

On other the hand,  BSs with a rotation were first studied in the work of Schunck and Mielke \cite{Schunck:1996}, 
and the rotating boson stars in four and five dimensions have been studied  in \cite{Kan:2016xkn}. After that, Yoshida and Eriguchi constructed the highly relativistic spinning BSs
\cite{1997PhRvD..56..762Y}. Moreover, the study of the spinning BSs solutions  can be extended to the quantization condition case \cite{Dias:2011at,Smolic:2015txa},  the quartic self-interacting potential as well as the Kerr black hole limit  case \cite{Herdeiro:2016gxs}. The linear stability of boson stars  with respect to small oscillations was discussed by Lee and Pang in \cite{Lee:1988av}; the study of the stability of boson stars was extended to the quartic and sextic self-interaction term case \cite{Kleihaus:2011sx} and  nonrotating  multistate boson stars \cite{Bernal:2009zy}; and the catastrophe theory  was applied to extract the stable branches of families of boson stars in \cite{Kusmartsev:2008py,Kusmartsev:1992}.

Recently, a class of Kerr black holes with a scalar hair was discussed by Herdeiro and Radu \cite{Herdeiro:2015gia,Herdeiro:2014goa}. The stability of a Kerr black hole with a scalar hair can be found in  Refs. \cite{Ganchev:2017uuo,Degollado:2018ypf,Hod:2012px,Benone:2014ssa}. In Refs. \cite{Herdeiro:2016tmi,Delgado:2016jxq,Herdeiro:2018wvd,Herdeiro:2018daq} the cases of the Proca hair, the Kerr-Newman black hole, nonminimal coupling case, and spinning black holes with a Skyrme hair have been achieved.
The study on long-term numerical evolutions of the superradiant instability of a Kerr black hole by East and Pretorius is found in  Ref. \cite{East:2017ovw}.
For a deeper analysis of the numerical methods and a review, see Refs.\cite{Herdeiro:2015gia,Herdeiro:2015waa}.
The family of a rotating Kerr black hole with a synchronised hair exhibits, besides  the physical quantities of mass and angular momentum, a conserved Noether charge  Q, which is associated with the complex scalar field, $\psi_{n}\sim e^{-iwt+im\varphi}$ (where $n\in \mathbb{N}_0$, $m\in \mathbb{Z}^*$) where the node number $n$ and  azimuthal harmonic index  $m$,
most of the studies of the solutions of Kerr black holes with scalar hair focused on the model of ground state ($n=0$) and  the smallest azimuthal harmonic index ($m=1$). Very recently, a family of the  Kerr black holes with an excited state scalar hair ($n\neq0$) have also been constructed \cite{Wang:2018xhw}, and  the Kerr black holes with odd parity scalar hair case was considered in \cite{Kunz:2019bhm} in detail.
The case of  Kerr black holes with synchronized hair and higher azimuthal harmonic index ($m>1$) have also  been investigated in Ref. \cite{Delgado:2019prc}.
In addition, the study of the spinning boson stars and hairy black holes is extended to a two-component Friedberg-Lee-Sirlin model  coupled to Einstein gravity in four spacetime dimensions \cite{Kunz:2019sgn}.
In the present work, we are interested in rotating multistate boson stars.  We  would like to know whether or not two scalar hairs can occupy the same state; furthermore,  we will construct  possible  coexisting states, including the ground and first excited states.

The paper is organized as follows.
In Sec. \ref{sec2}, we introduce  the model of
the four-dimensional Einstein gravity coupled to two complex massive scalar fields $\psi_i$ $(i=1,2)$  and adopt the same axisymmetric metric with Kerr-like coordinates as the  ansatz in Ref. \cite{Herdeiro:2014goa}.
In Sec. \ref{sec3}, the boundary conditions of the rotating multistate boson stars (RMSBS) are studied.
We show the numerical  results of the equations of motion and show the characteristics of the $^1S^2S$ state and the $^1S^2P$ state in Sec. \ref{sec4}. We conclude in Sec. \ref{sec5} with a discussion and an outline for further work.

\section{The model setup}\label{sec2}
We start with the theory of  Einstein gravity  coupled to two massive complex  scalar fields $\psi_i$ $(i=1,2)$,
\begin{align}
S=\int_{\mathcal{M}} \mathrm{d}^4x\sqrt{-g}\left(\frac{R}{16\,\pi\,G}-\nabla_a\psi_1^*\nabla^a\psi_1-\mu_1^2|\psi_1|^2-\nabla_a\psi_2^*\nabla^a\psi_2-\mu_2^2|\psi_2|^2\right)\,,
\end{align}
where $\mu_i$ $(i=1,2)$ are the mass of two scalar fields, respectively. From  henceforth, we will set $G=c=1$. The corresponding equations of motion are given by
\begin{subequations}
\begin{equation}
\frac{R_{ab}}{8\pi}=2\nabla_{(a}\psi_1^*\nabla_{b)}\psi_1+g_{ab}\mu_1^2\psi_1^*\psi_1+2\nabla_{(a}\psi_2^*\nabla_{b)}\psi_2+g_{ab}\mu_2^2\psi_2^*\psi_2\,,
\label{eq:EKG1}
\end{equation}
\begin{equation}
\Box\psi_1=\mu_1^2\psi_1\,,
\label{eq:EKG2}
\end{equation}
\begin{equation}
\Box\psi_2=\mu_2^2\psi_2\,.
\label{eq:EKG3}
\end{equation}

\label{eq:EKG}%
\end{subequations}
When both of the two scalar fields vanish, the solution of  Eq.
(\ref{eq:EKG1}) has the stationary axisymmetric asymptotically flat black hole with a mass
and angular momentum, which is the well-known Kerr black hole. In terms of Boyer-Lindquist
coordinates, the Kerr metric reads
\begin{eqnarray}
\mathrm{d}s^2 = -\frac{\Delta}{\Sigma^2}\left(\mathrm{d}t-a \sin^2\theta \mathrm{d}\phi\right)^2+\frac{\sin^2\theta}{\Sigma^2}[a\,\mathrm{d}t-(r^2+a^2)\mathrm{d}\phi]^2
+\Sigma^2\left(\mathrm{d}\theta^2+\frac{\mathrm{d}r^2}{\Delta}\right)\,,
\label{eq:Boson star}
\end{eqnarray}
with $\Delta = r^2+a^2-2\,M\,r$ and $\Sigma^2=r^2+a^2\cos^2\theta$. The black hole event horizon is a null hypersurface with $r=r_+\equiv M+\sqrt{M^2-a^2}$, angular velocity $\Omega_K = a/(a^2+r_+^2)$, and temperature $T_K=(r_+^2-a^2)/[4\pi r_+(r_+^2+a^2)]$. The constant $M$ is the black hole mass and $a$ parameterized is the angular momentum via $J = M\,a$.

In Refs. \cite{Herdeiro:2014goa, Herdeiro:2015gia}, Herdeiro and Radu constructed a family of boson stars as well as a Kerr black hole with a ground state scalar hair. In order to construct stationary solutions of the RMSBS, we also take the same numerical method with   the following ansatz:
\begin{eqnarray}
\label{ansatz}
ds^2=e^{2F_1}\left(\frac{dr^2}{N }+r^2 d\theta^2\right)+e^{2F_2}r^2 \sin^2\theta (d\varphi-W dt)^2-e^{2F_0} N dt^2,
\end{eqnarray}
with $N=1-\frac{r_H}{r}\ ,$ and the constant $r_H$ that is related to event horizon radius. Besides, the ansatz of  two complex scalar fields $\psi_i$ are given by
\begin{eqnarray}
\psi_i=\phi_{i(n)}(r,\theta)e^{i(m_i\varphi-\omega_i t)}, \;\;\;  i=1,2,\;\;\; n=0,1,\cdots,\;\;\; m_i=\pm1,\pm2,  \cdots.
\label{scalar_ansatz1}
\end{eqnarray}
Here, we note that the six functions $F_0$, $F_1$, $F_2$, $W$, and  $\phi_{i(n)}$ $(i=1,2)$ depend on the radial distance $r$ and polar angle $\theta$.  Again, the constants $\omega_i$ $(i=1,2)$ are the frequency of the complex scalar field and $m_i$ $(i=1,2)$ are the azimuthal harmonic index, respectively.
When  $\omega_1=\omega_2=\omega$, the frequency of the scalar field is called the  synchronized frequency, while $\omega_1\neq\omega_2$ is  called the nonsynchronized frequency. The subscript  $n$  of Eq.(\ref{scalar_ansatz1}) is  named as the principal quantum number of the scalar field, and  $n=0$ is regarded as the ground state and $n\geq1$ as the excited states.
Besides, in the scalar ansatz (\ref{scalar_ansatz1}), subscripts $i$  are indicated by two complex  scalar fields only.

It is well known that the ground state scalar hair has no node,  that is, along the radial $r$ direction, the value of the scalar field has the same sign.
For the rotating boson stars with  first excited state, we observe that there are two types of nodes,  radial and angular nodes.
Radial nodes are the points where the value of  the scalar field can  change sign along the radial $r$ direction, while  angular nodes  are the points where the value of the scalar field can  change sign along the angular $\theta$ direction.
Hence,  we would like to construct rotating boson stars composed of  two coexisting states of the  scalar fields, including the ground state and the first excited state.

\section{Boundary conditions}\label{sec3}

Before  numerically solving the differential equations instead of seeking the analytical solutions, we should obtain the asymptotic behaviors of the  metric functions $F_{0}(r,\theta),F_{1}(r,\theta),F_{2}(r,\theta)$, and  $W(r,\theta)$  as well as the scalar field $\phi_{i(n)}(r,\theta)$ $(i=1, 2)$,  which is equivalent to knowing the boundary conditions we need.
Considering  the  properties of the RMSBS, we will still use  the  boundary conditions by following the same steps as  given in Refs. \cite{Herdeiro:2014goa,Herdeiro:2015gia,Wang:2018xhw}.

For rotating axially symmetric boson stars, exploiting the reflection symmetry  $\theta\rightarrow\pi-\theta$ on the equatorial plane,  it is enough to consider the  range $\theta \in [0,\pi/2] $ for the angular variable.
At infinity $r\rightarrow\infty$,  the boundary conditions  are
\begin{equation}\label{rbc}
  F_0=F_1=F_2=W=\phi_{i(n)}=0,\;\;\; (i=1, 2), \;\;\;  n=0,1,\cdots,
\end{equation}
and we require the boundary conditions,
\begin{equation}\label{abc}
\partial_\theta F_0(r, 0)=\partial_\theta F_1(r, 0)=\partial_\theta F_2(r, 0) =\partial_\theta W(r, 0)=\phi_{i(n)}(r, 0)=0,\;\;\;n=0,1,\cdots,
\end{equation}
 for $\theta=0$.
For odd parity solutions, we have
\begin{equation}
\partial_\theta F_0(r,\pi/2)=\partial_\theta F_1(r,\pi/2)=\partial_\theta F_2(r,\pi/2) = \partial_\theta W(r,\pi/2)= \phi_{i(n)}(r,\pi/2) = 0,\;\;\; n=1,2,\cdots,
\end{equation}
 for $\theta=\pi/2$,
while for even parity solutions, $\partial_\theta \phi_{i(n)}(r, \pi/2) = 0$ with $ n=1,2,\cdots$.

For rotating multistate boson stars solutions with $r_H=0$,
\begin{eqnarray}
 \phi_{i(n)}(0, \theta) = 0,  \nonumber\\
 \partial_r W(0, \theta) = 0.
\end{eqnarray}
We note that the values of $F_0(0, \theta), F_1(0, \theta), F_2(0, \theta)$, and  $W(0,\theta)$ are the constants that are not dependent of the polar angle $\theta$.

Near the boundary $r\rightarrow\infty$, on the other hand, the mass of boson stars $M$  and the total angular momentum  $J$  are extracted from the  asymptotic behavior of  the metric functions,
\begin{eqnarray}
\label{asym}
g_{tt}= -1+\frac{2GM}{r}+\cdots, \nonumber\\
g_{\varphi t}= -\frac{2GJ}{r}\sin^2\theta+ \cdots.
\end{eqnarray}

\section{Numerical results}\label{sec4}
In this section, we will solve  the above coupled Eqs. (\ref{eq:EKG1}), (\ref{eq:EKG2}), and  (\ref{eq:EKG3}) with the ansatzs (\ref{ansatz}) and (\ref{scalar_ansatz1}) numerically;
it is convenient to change  the    radial coordinate $r$ to
\begin{eqnarray}
 x =\frac{\sqrt{r^2-r_H^2}}{1+\sqrt{r^2-r_H^2}},
\end{eqnarray}
which implies that the new radial coordinate $x  \in [0,1]$.  Thus, the inner and outer boundaries of the shell
are fixed at $x = 0$ and  $x = 1$, respectively.  By exploiting the reflection symmetry  $\theta\rightarrow\pi-\theta$ on the equatorial plane,  it is enough to consider the  range $\theta \in [0,\pi/2] $ for the angular variable.
In addition, there exist two classes of solutions: horizonless boson star solutions with $r_{H}=0$ and hairy black hole solutions with $r_{H}>0$.
In this paper, however, we mainly consider the boson star solutions with $r_{H}=0$.

Before numerically solving the equations, we can study the dependence on the synchronized frequency $\omega$, the nonsynchronized frequencies $\omega_1$, $\omega_2$, and the scalar field masses $\mu_1$ and $\mu_2$, respectively. To simplify our analysis, we can work at a fixed value of only one of the scalar field masses; for instance, $\mu_1=1$.

All numerical calculations are based on the finite element methods.
Typical grids used have sizes of $100\times100$ in the integration region $0\leq x\leq 1$ and $0\leq\theta\leq\frac{\pi}{2}$. Our iterative process is the Newton-Raphson method, and the relative error for the numerical solutions in this work is estimated to be below $10^{-5}$.

Next,  we will discuss the RMSBS, including the principal quantum number $n=0$, which is the ground state case and
the principal quantum number $n=1$, which belongs to the case of the first excited state. Besides, we exhibit two classes of  radial $n_r=1$  and angular  $n_\theta=1$ node solutions, respectively.
As noted above, for the case of  rotating boson stars with the first excited state, there is a similar situation in atomic theory and quantum mechanics; the first excited state of hydrogen has an electron in the 2s orbital and the 2p orbital, which correspond to the radial and angle node, respectively. Therefore, the coexisting states of two  scalar fields, which have a ground state and a first excited state with a radial node $n_r=1$, is  named  as the $^1S^2S$ state. Besides, the  coexistence of   a ground state and a first excited state with an angle node $n_\theta=1$ is called the $^1S^2P$ state.

\subsection{ $^1S^2S$ state}
In this subsection, we will study the solutions with an even-parity scalar field.  Along the angular $\theta$ direction, the values of the scalar fields $\phi_1$ and $\phi_2$ have the same sign.  Along the radial $r$ direction,  the  scalar field  $\phi_1$  keeps the  same sign, while,
 the  scalar field  $\phi_2$  changes sign once at some point.
 From the view of the excited states, these two states are just similar to the 1-s  and 2-s states of the hydrogen atom, respectively.
\subsubsection{Boson star}
\begin{figure}[t]
\centering
  \begin{minipage}[t]{0.3\textwidth}
    \includegraphics[width=\textwidth]{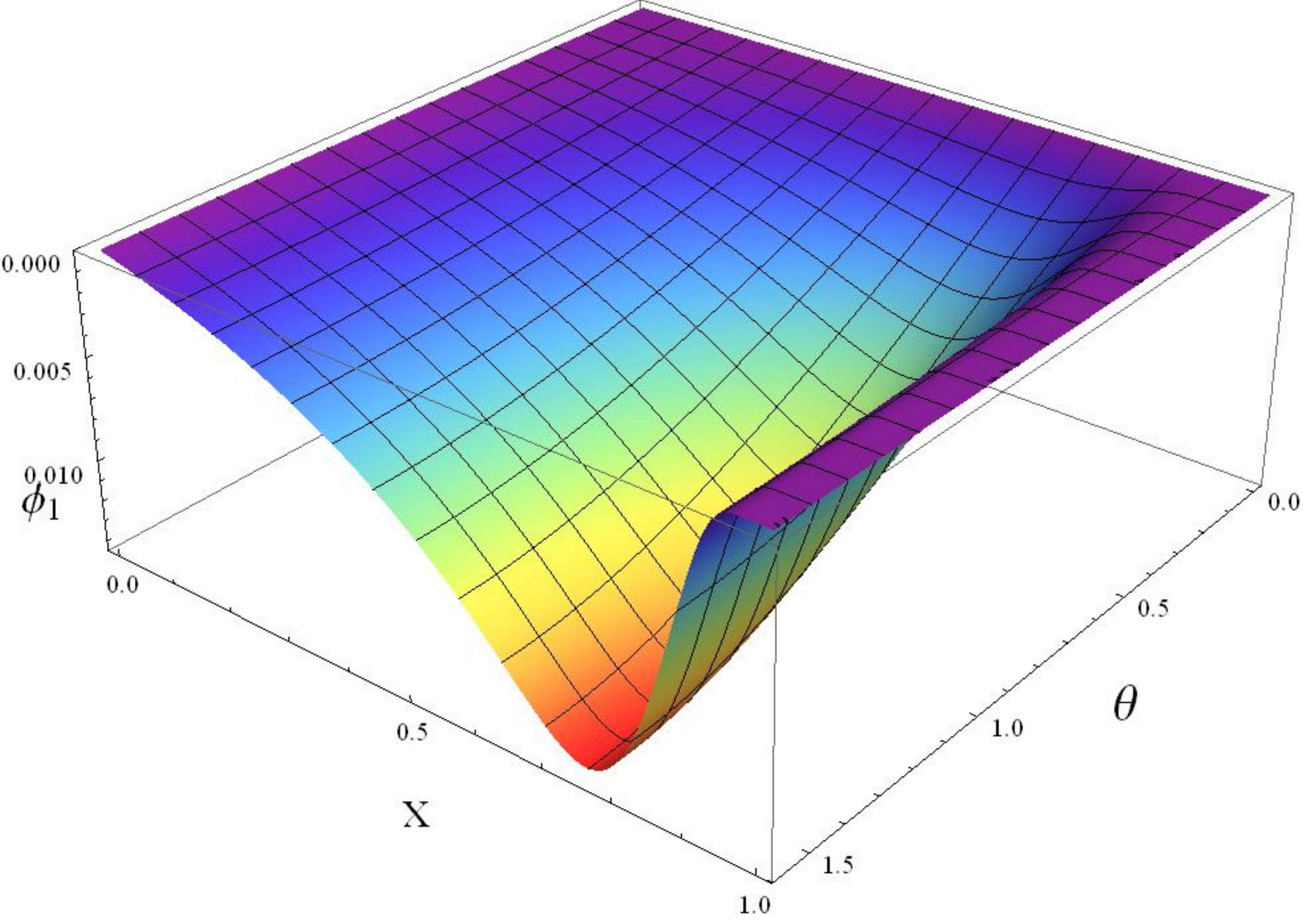}
  \end{minipage}
    \hfill
    \begin{minipage}[t]{0.3\textwidth}
    \includegraphics[width=\textwidth]{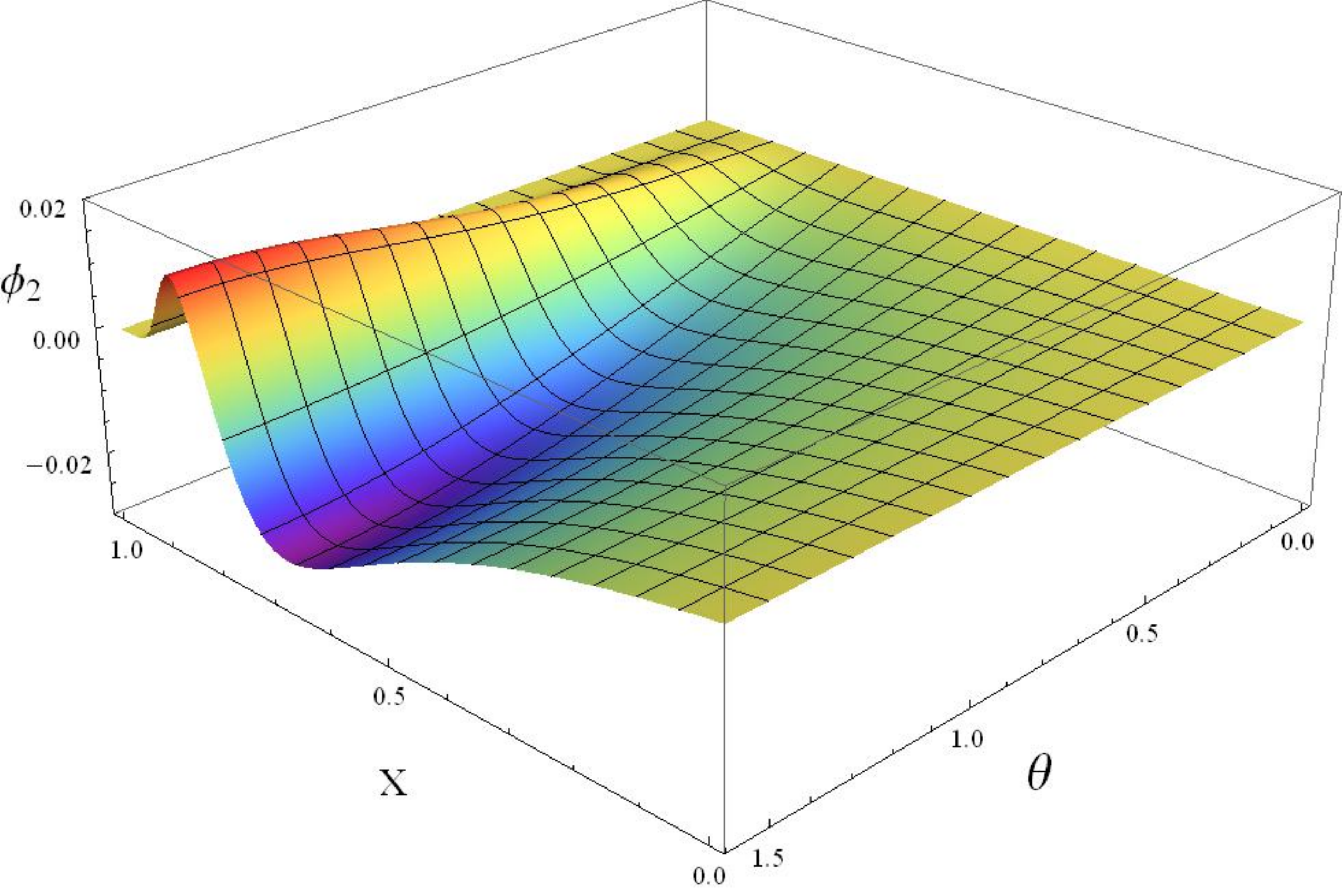}
      \end{minipage}
     \hfill
     \begin{minipage}[t]{0.32\textwidth}
    \includegraphics[width=\textwidth]{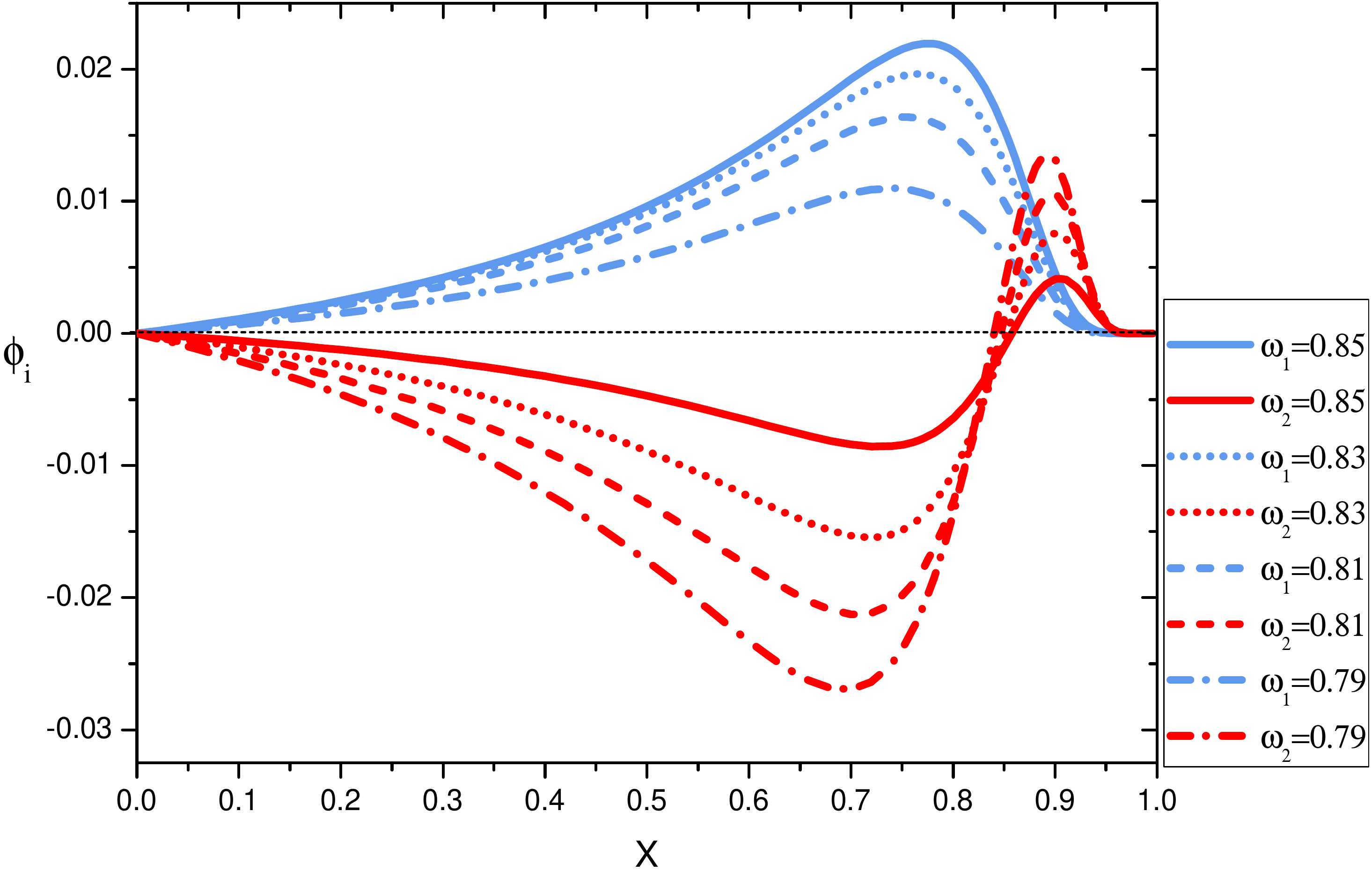}
  \end{minipage}
   \caption{ The distribution of the scalar field $\phi_1$ as a function of $x$ and $\theta$ (left panel) and  the scalar $\phi_2$ as a function of $x$ and $\theta$ (middle panel) with
   the same parameter $\omega_1=\omega_2=0.8$, as well as the  numerical solutions of the scalar fields $\phi_i$  $(i=1,2)$  versus the boundary $x$, represented by the blue and red lines, respectively and the horizon dashed black line represents the  zero value (right panel). All solutions have $m_1=m_2=1$, $\mu_1=1$, and $\mu_2=0.93$.}
  \label{fig:f-11}
\end{figure}
\begin{figure}[t]
\centering
  \begin{minipage}[t]{0.48\textwidth}
    \includegraphics[width=\textwidth]{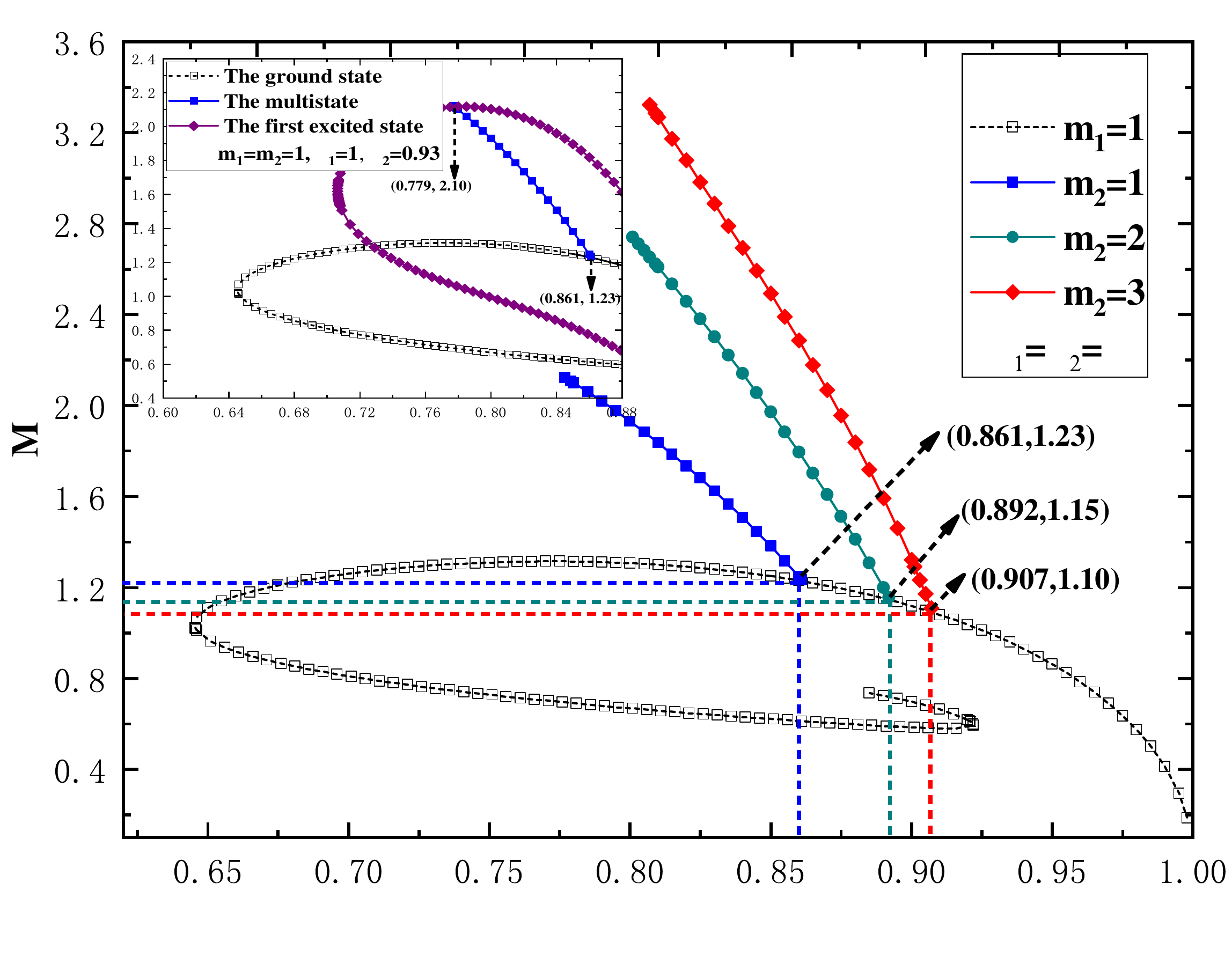}
  \end{minipage}
    \hfill
    \begin{minipage}[t]{0.48\textwidth}
    \includegraphics[width=\textwidth]{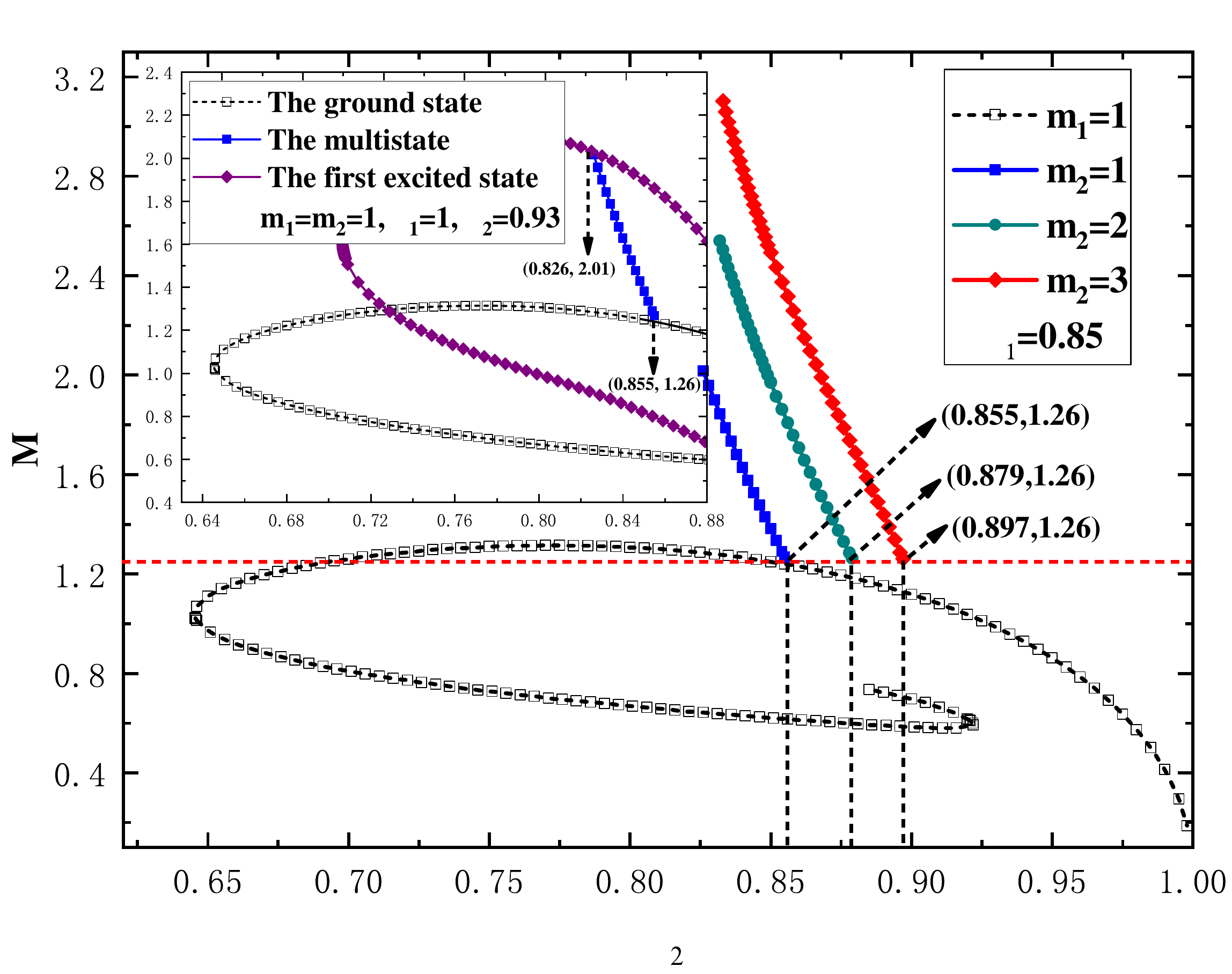}
  \end{minipage}
   \caption{\textit{Left}: The mass $M$ of the RMSBS as a function of the synchronized frequency $\omega$ with an azimuthal harmonic index $m_{2}=1~,2,~3$. Three intersection points correspond to the coordinates $(0.861,~1.23)$, $(0.892,~1.15)$, and $(0.907,~1.10)$, respectively.  \textit{Right}: The mass $M$ of the  RMSBS as a function of the nonsynchronized frequency $\omega_2$ with the fixed parameter $\omega_1=0.85$. The horizon red dashed line indicate the mass $M=1.26$,  and the right ends  of the blue,  cyan, and red dotted lines correspond to the  same value of  $M$ with coordinates  $(0.855,~1.26)$,  $(0.879,~1.26)$, and $(0.897,~1.26)$, respectively. In both panels, the black hollowed line indicates the ground state solutions, and  the inset of  both panels show the ground state with $\mu_1=1$ (black line),  the first excited state with $\mu_2=0.93$ (purple line), and the multistate with $\mu_1=1,\mu_2=0.93$ (blue line). All solutions have $m_1=1$.}
  \label{fig:f-12}
\end{figure}

Numerical results are presented in Fig.~\ref{fig:f-11}. We present the scalar field $\phi_1$ (left panel) and $\phi_2$  (middle panel) as a function of $x$ and $\theta$
with the azimuthal harmonic index $m_1=m_2=1$ for the same frequency $\omega_1=\omega_2=0.8$.
The distribution of the scalar field $\phi_1$ (blue lines) and the scalar field $\phi_2$ (red lines)
versus the boundary $x$ for several values of frequency $\omega_1=\omega_2$ are exhibited in the right panel of Fig.~\ref{fig:f-11};
we can observe that the scalar field $\phi_2$ changes sign once from the center of the boson stars to the boundary in a node.
These behaviors are further shown in Fig.~\ref{fig:f-12}.

\begin{figure}[t]
\centering
  \begin{minipage}[t]{0.48\textwidth}
    \includegraphics[width=\textwidth]{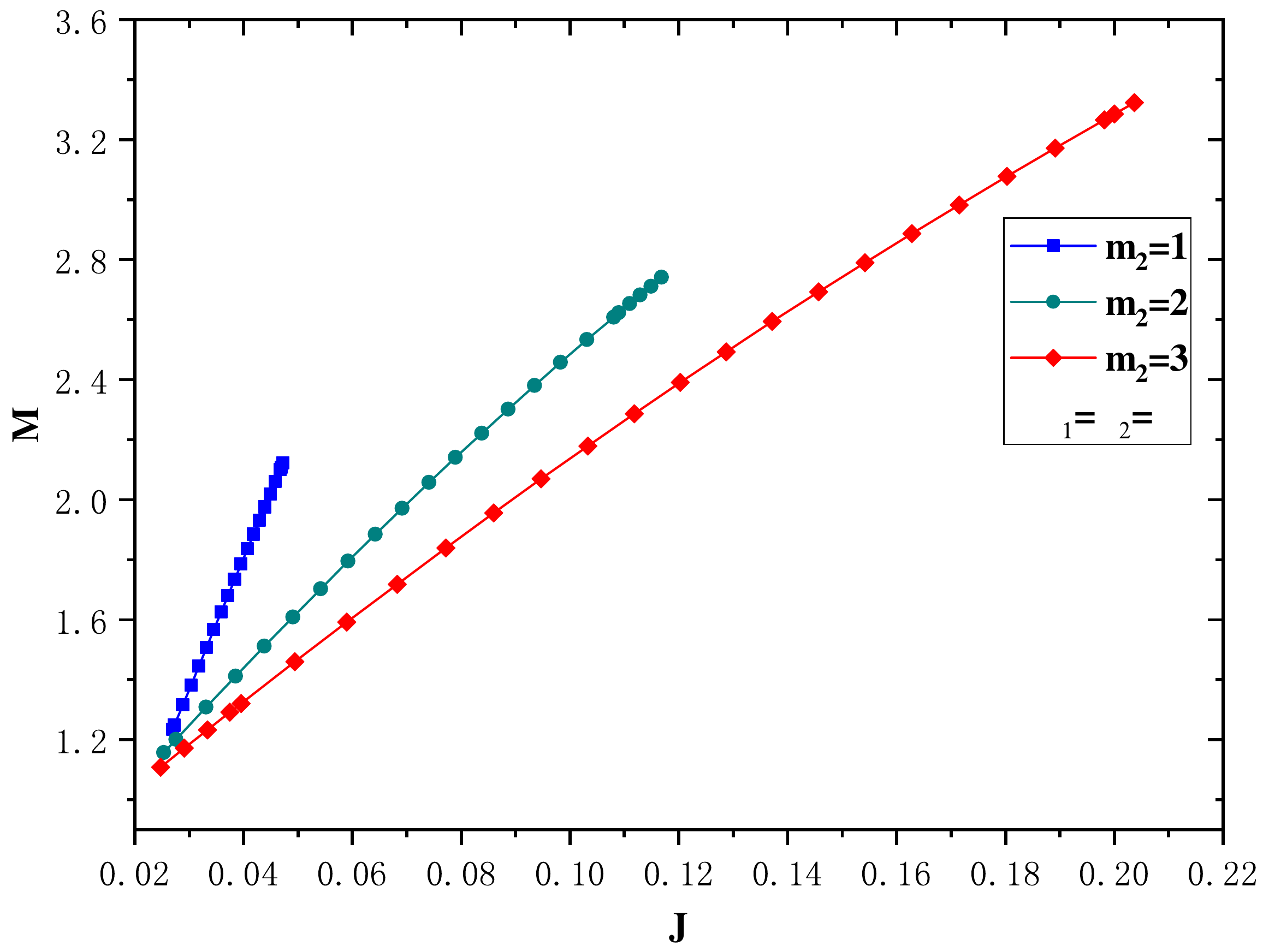}
  \end{minipage}
    \hfill
    \begin{minipage}[t]{0.48\textwidth}
    \includegraphics[width=\textwidth]{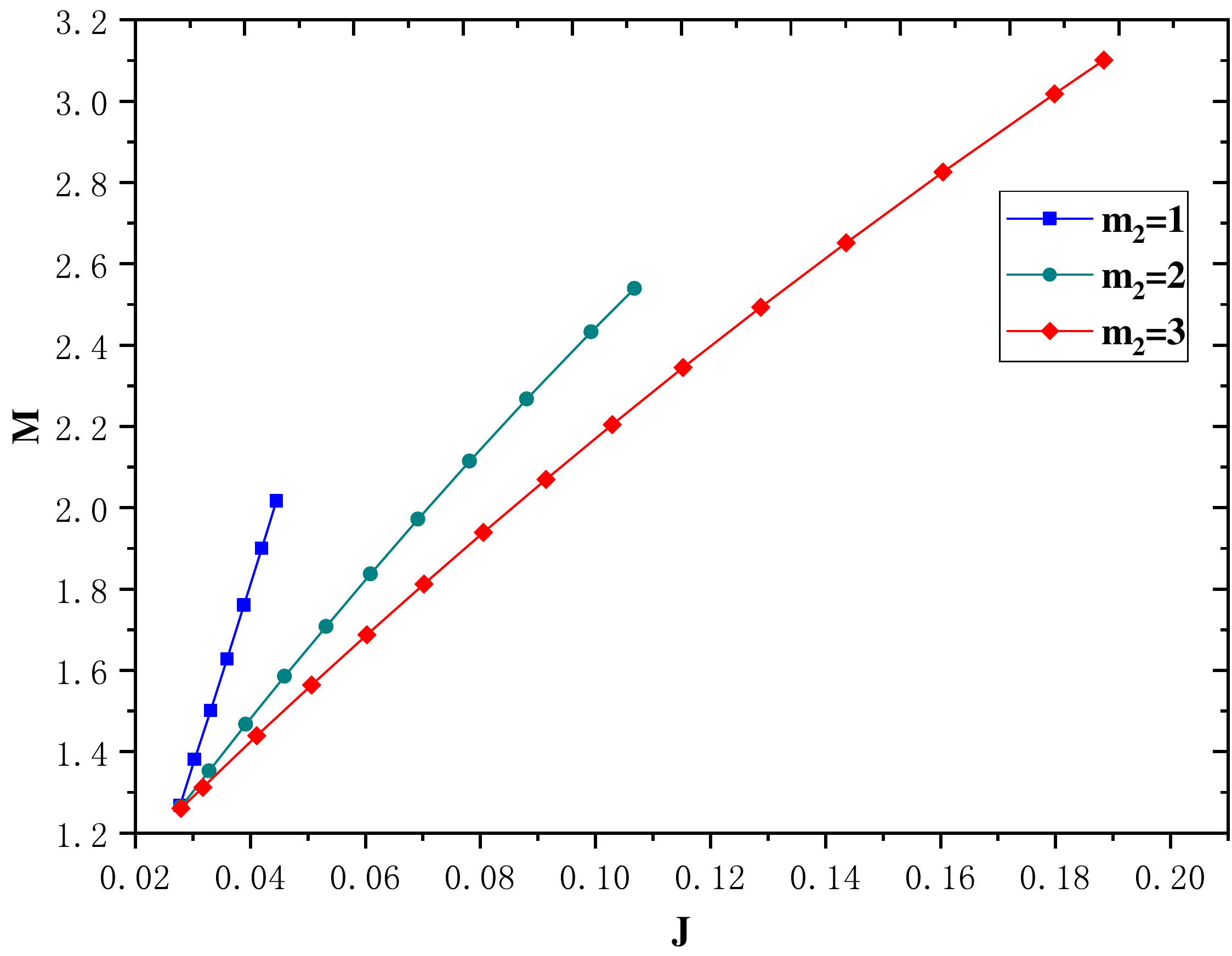}
  \end{minipage}
   \caption{\textit{Left}: The mass $M$ of  the RMSBS  versus the angular momentum $J$ for the synchronized frequency $\omega$ with $m_{2}=1~,2,~3$ , respectively. \textit{Right}: The mass $M$ of the RMSBS versus the angular momentum $J$ for the nonsynchronized frequency $\omega_2$ with $m_{2}=1~,2,~3$, respectively, and we demand the parameter $\omega_1=0.85$.}
  \label{fig:M-J-1}
\end{figure}

Meanwhile, to discuss the properties of  the RMSBS, we  also simplify our analysis; we mainly exhibit in  Figs.~\ref{fig:f-12} and ~\ref{fig:M-J-1} the mass $M$ and the angular momentum $J$ of several sets of the RMSBS versus the synchronized frequency $\omega$ and the nonsynchronized frequency $\omega_2$ with  the azimuthal harmonic index $m_{2}=1~,2,~3$.

The left panel of Fig.~\ref{fig:f-12} exhibits the variation of the mass of the RMSBS versus the synchronized frequency $\omega$ with the azimuthal harmonic index $m_{2}=1~,2,~3$, represented by the blue, cyan, and red lines, respectively, and the black hollowed line indicates the ground state with $m_1=1$.
First observe that the  domain of existence of the RMSBS  are similar to the ground state boson stars in Ref. \cite{Herdeiro:2014goa}.
We again observe that as the synchronized frequency $\omega$ decreases, the mass of the RMSBS keeps increasing.
In Ref. \cite{Herdeiro:2014goa}, the behavior of the ground state solutions with $m_1=1$ spirals to the center;
however, the RMSBS case does not occur with a second branch of unstable solutions with $m_{2}=1~,2,~3$.
In addition, we observe that,  as the azimuthal harmonic index $m_2$ increases, the maximum value of the synchronized frequency $\omega$ also increases, and the minimum value of the mass of the RMSBS  decreases as $m_2$ increases, and the multistate curves for $m_{2}=2,~3$ intersect with the ground state solutions with  the coordinates  $(0.892,~1.15)$ and $(0.907,~1.10)$, respectively.
From the inset in left panel of Fig.~\ref{fig:f-12}, we can see that the curves of the ground state with $\mu_1=1$ (black), and first excited states with $\mu_2=0.93$ (purple). The multistate curve with $\mu_1=1,\mu_2=0.93$ (blue)  intersects with the ground  and first excited states with  the coordinates $(0.861,~1.23)$ and $(0.779,~2.10)$, respectively.
This means that when the synchronized frequency $\omega$ tends to its maximum, the first excited state could reduce to zero and there exists only a
single the ground state. On the contrary, with the decrease of the synchronized frequency $\omega$, there exists only  the first excited  state.

In the right panel of Fig.~\ref{fig:f-12}, we plot the mass of the RMSBS versus the nonsynchronized frequency $\omega_2$ for the fixed value of $\omega_1=0.85$.
One observes that, by increasing the azimuthal harmonic index $m_2$, the mass of the RMSBS keeps increasing. Meanwhile, as $\omega_2$ increases to its maximum,
the minimum value of the mass of the RMSBS is the constant value $M=1.26$; three coordinates correspond to $(0.855,~1.26)$, $(0.879,~1.26)$, and $(0.897,~1.26)$, respectively.
From the inset in right panel of Fig.~\ref{fig:f-12}, we show  the ground state with $\mu_1=1$ (the black lines) and first excited states with $\mu_2=0.93$ (the purple lines). The multistate with $\mu_1=1,\mu_2=0.93$ (the blue line) intersects with the ground  and first excited states with  coordinates  $(0.855,~1.26)$ and   $(0.826,~2.01)$, respectively.
That is, as the  nonsynchronized frequency $\omega_2$ increases,
the first excited state could decrease to zero and the mass of the RMSBS is provided by the ground state. On the contrary, with the decrease of the  nonsynchronized frequency $\omega_2$, there exists only  the first excited  state.
While we fixed the value of $\omega_1=0.85$,  the minimal mass of the RMSBS is always a constant value $M=1.26$ for the different azimuthal harmonic indexes $m_2=2,~3$.

\begin{figure}[t]
\centering
  \begin{minipage}[t]{0.52\textwidth}
    \includegraphics[width=\textwidth]{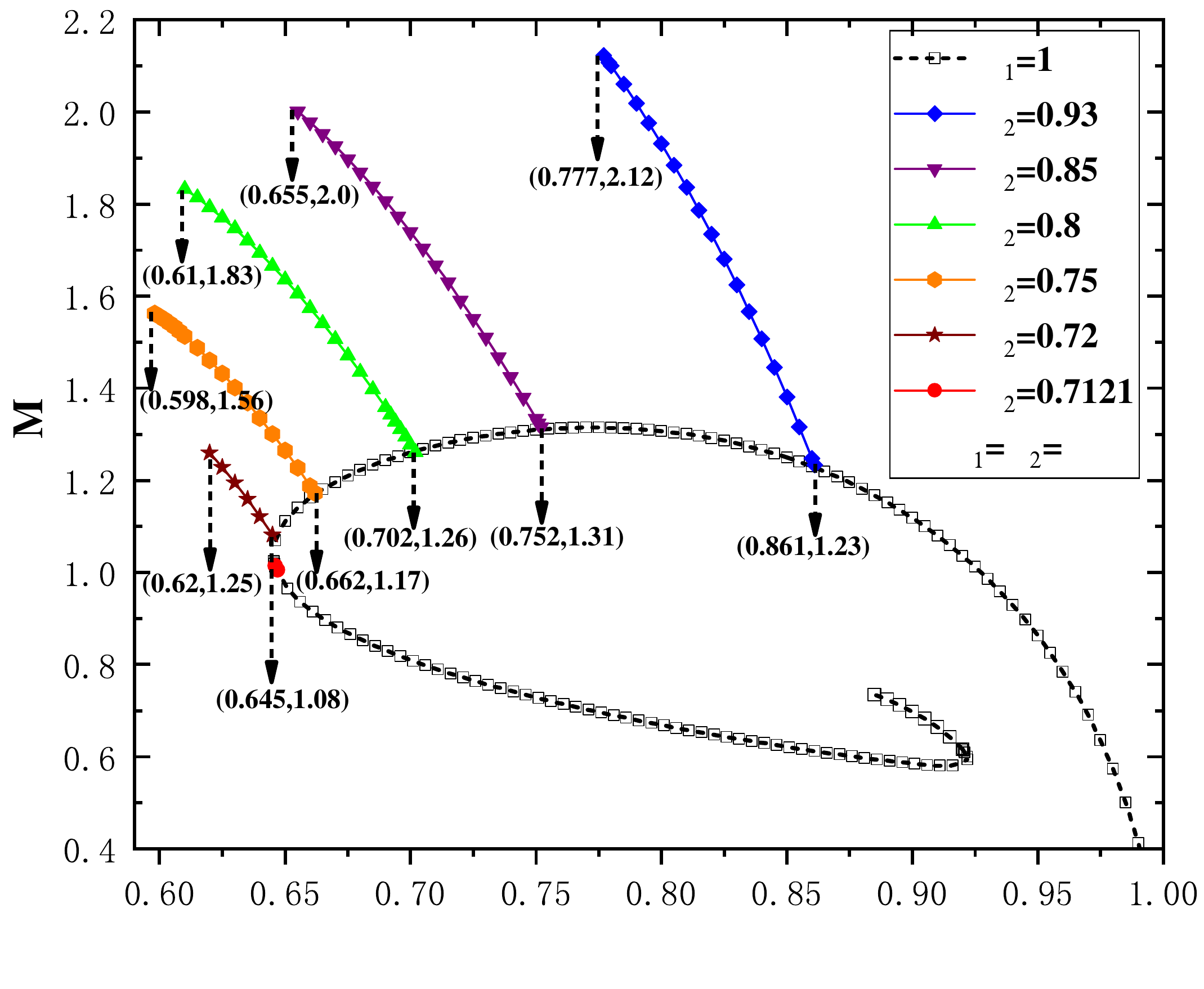}
  \end{minipage}
   \caption{ The mass $M$ of  RMSBS as a function of synchronized frequency $\omega$ for the different values of $\mu_2$. The black hollowed line denotes the ground state with $\mu_1=1$, and all solutions have $m_{1}=m_{2}=1$ and $\mu_1=1$. }
  \label{fig:f-13}
\end{figure}

In order to verify whether there exists another family of multistate solutions  between the ground state and the first excited state, we  use  two  methods to seek for the new family of multistate solutions. The first way is that we adopt the similar method as the numerical algorithm given in Sec.VIIB of \cite{Dias:2015nua}, and the second way is at the fixed $\mu_1=1$ where we change the values of $\mu_2$ to seek for the new family of multistate solutions for the same parameters. However, we fail to find the other family of multistate solutions in the maximum and minimum of the frequency for the cases of the $^1S^2S$ and $^1S^2P$ states. As an example in Fig.~\ref{fig:f-13}, the mass $M$ of RMSBS is a function of synchronized frequency $\omega$ for the different values of $\mu_2$. We observe that the domain of existence of the synchronized frequency $\omega$ decreases with the decrease of the scalar field mass $\mu_2$. Furthermore, when $\mu_2\rightarrow0.7121$, the multistate curves approach the turning point of the ground state curves, and the synchronized frequency $\omega$ have a narrower range with $(0.646,~0.647)$. However, we fail to find the new family of multistate solutions between the ground state and the first excited state, but that does not mean there does not exist any new family of multistate solutions. At present, it is difficult for us to find out the other family of multistate solutions with numerical methods.

Moving on with our analysis, we now consider the variation of the solutions with the  mass  $M$ of the rotating multistate boson stars  that varies versus the angular momentum $J$, which is dependent on the frequency.
In Fig.~\ref{fig:M-J-1} (left panel), we exhibit the  mass $M$  of  the RMSBS  versus the angular momentum $J$  with  different azimuthal harmonic index  $m_2 = 1$ (blue lines),  2 (cyan lines), and   3 (red lines) for the synchronized frequency $\omega$. In the right panel of Fig.~\ref{fig:M-J-1}, the mass $M$  versus the angular momentum $J$  with  different azimuthal harmonic indexes $m_2=1~,2,~3$ for the nonsynchronized frequency $\omega_2$ are shown, and we set the frequency parameter $\omega_1=0.85$. Comparing with the results of the ground state boson stars in Ref. \cite{Herdeiro:2014goa}, we can see that the case of the RMSBS does not occur in zigzag patterns, and the minimum value of the mass $M$ is larger than that of the ground state boson stars.
\begin{table*}[htp]
\caption{The domain of existence of the  mass $\mu_2$ of the scalar field $\phi_2$ in  three different situations: The synchronized frequency $\omega_1=\omega_2=\omega$ (left panel), the nonsynchronized frequency $\omega_1$ (middle panel), and  the nonsynchronized frequency $\omega_2$ (right panel) with $m_2=1,2,3$, respectively. In the middle and right panels, we adopt  $\omega_2=0.85$ and $\omega_1=0.85$, respectively. All solutions have $\mu_1=1$ and $m_1=1$.}
\centering
\subtable[
]{
\scalebox{0.62}{
       \begin{tabular}{|c|c|c|c|}
       \midrule[1.0pt]\midrule[1.0pt]
\diagbox{$\omega_{~}$}{$\mu_{2}$}{$m_{2}$}&$1$&$2$&$3$\\ \midrule[1.0pt]
$0.78$&$0.874\sim0.931$&$0.828\sim0.920$&$0.803\sim0.917$\\
\hline
$0.81$&$0.897\sim0.944$&$0.859\sim0.934$&$0.836\sim0.931$\\
\hline
$0.84$&$0.917\sim0.955$&$0.887\sim0.947$&$0.868\sim0.944$\\
\hline
$0.86$&$0.930\sim0.962$&$0.905\sim0.955$&$0.887\sim0.953$\\
\hline
$0.88$&$0.942\sim0.969$&$0.921\sim0.963$&$0.906\sim0.961$\\
\midrule[1.0pt]\midrule[1.0pt]
\end{tabular}}
       \label{tab:firsttable}
}
\subtable[
]{
\scalebox{0.615}{
       \begin{tabular}{|c|c|c|c|}
       \midrule[1.0pt]\midrule[1.0pt]
\diagbox{$\omega_{1}$}{$\mu_{2}$}{$m_{2}$}&$1$&$2$&$3$\\ \midrule[1.0pt]
$0.70$&$1.058\sim1.120$&$0.927\sim1.080$&$0.879\sim1.070$\\
\hline
$0.80$&$0.956\sim1.004$&$0.910\sim0.991$&$0.882\sim0.986$\\
\hline
$0.84$&$0.930\sim0.968$&$0.899\sim0.959$&$0.878\sim0.956$\\
\hline
$0.86$&$0.918\sim0.951$&$0.893\sim0.944$&$0.876\sim0.942$\\
\hline
$0.88$&$0.907\sim0.934$&$0.887\sim0.929$&$0.873\sim0.927$\\
\midrule[1.0pt]\midrule[1.0pt]
\end{tabular}}
       \label{tab:secondtable}
}
\subtable[]{
\scalebox{0.615}{
       \begin{tabular}{|c|c|c|c|}
       \midrule[1.0pt]\midrule[1.0pt]
\diagbox{$\omega_{2}$}{$\mu_{2}$}{$m_{2}$}&$1$&$2$&$3$\\ \midrule[1.0pt]
$0.83$&$0.899\sim0.935$&$0.873\sim0.928$&$0.855\sim0.926$\\
\hline
$0.84$&$0.911\sim0.947$&$0.884\sim0.940$&$0.866\sim0.937$\\
\hline
$0.85$&$0.924\sim0.959$&$0.896\sim0.951$&$0.878\sim0.949$\\
\hline
$0.86$&$0.936\sim0.971$&$0.908\sim0.963$&$0.889\sim0.960$\\
\hline
$0.88$&$0.959\sim0.994$&$0.930\sim0.985$&$0.911\sim0.983$\\
\midrule[1.0pt]\midrule[1.0pt]
\end{tabular}}
       \label{tab:thirdtable}
}
\label{tab:table1}
\end{table*}

In  Table \ref{tab:table1}, we show the domain of existence of the  mass $\mu_2$ of the scalar field $\phi_2$ in  three different situations. The mass $\mu_2$  versus the synchronized frequency $\omega$  in Table  \ref{tab:firsttable}, the nonsynchronized frequency $\omega_1$  in Table \ref{tab:secondtable}, and the nonsynchronized frequency $\omega_2$  in Table \ref{tab:thirdtable}, as well as the three subtables have the same azimuthal harmonic index parameters $m_2=1~,2,~3$, respectively.

In order to explore the influence of the different typical frequencies, the  domain of existence of the mass $\mu_2$ with the azimuthal harmonic index parameter $m_2=1~,2,~3$ for the same parameters $\omega=\omega_1=\omega_2=0.84~,0.86,~0.88$ are shown. From  Table \ref{tab:firsttable}, it is obvious that the domain of existence of the mass $\mu_2$ decreases with increasing synchronized frequency $\omega$. Again, by increasing the value of the azimuthal harmonic index parameter $m_2$, the mass domain as the synchronized frequency $\omega$ keeps increasing.
In order to compare  the results of the domain of existence of the mass $\mu_2$ of the scalar field $\phi_2$ versus the synchronized frequency $\omega$, we exhibit the domain of existence of the mass $\mu_2$ of the scalar field $\phi_2$ as a function of the nonsynchronized frequency $\omega_1$ in Table \ref{tab:secondtable} and $\omega_2$ in Table \ref{tab:thirdtable} for the azimuthal harmonic index parameters $m_2=1~,2,~3$.
On the other hand,  in Tables ~\ref{tab:secondtable} and ~\ref{tab:thirdtable},  the domain of existence of the mass $\mu_2$ with $m_2=1,~2,~3$ is similar to the case of  synchronized frequency $\omega$ in Table \ref{tab:firsttable}, respectively.

\subsection{ $^1S^2P$ state}
In  this subsection, we exhibit  the solutions with one even-parity and  one odd-parity scalar field. Along the angular $\theta$ and  the radial $r$ directions,
the  scalar fields  $\phi_1$ and $\phi_2$ also keep the  same sign.
Moreover, it is noted that the configuration with an odd parity are more unstable than the case with an even-parity scalar field  \cite{Herdeiro:2015gia,Wang:2018xhw}.
From the view of the excited states, these two states are just similar to the 1-s and  2-p states of the hydrogen atom, respectively.
\subsubsection{Boson star}
\begin{figure}[t]
\centering
  \begin{minipage}[t]{0.3\textwidth}
    \includegraphics[width=\textwidth]{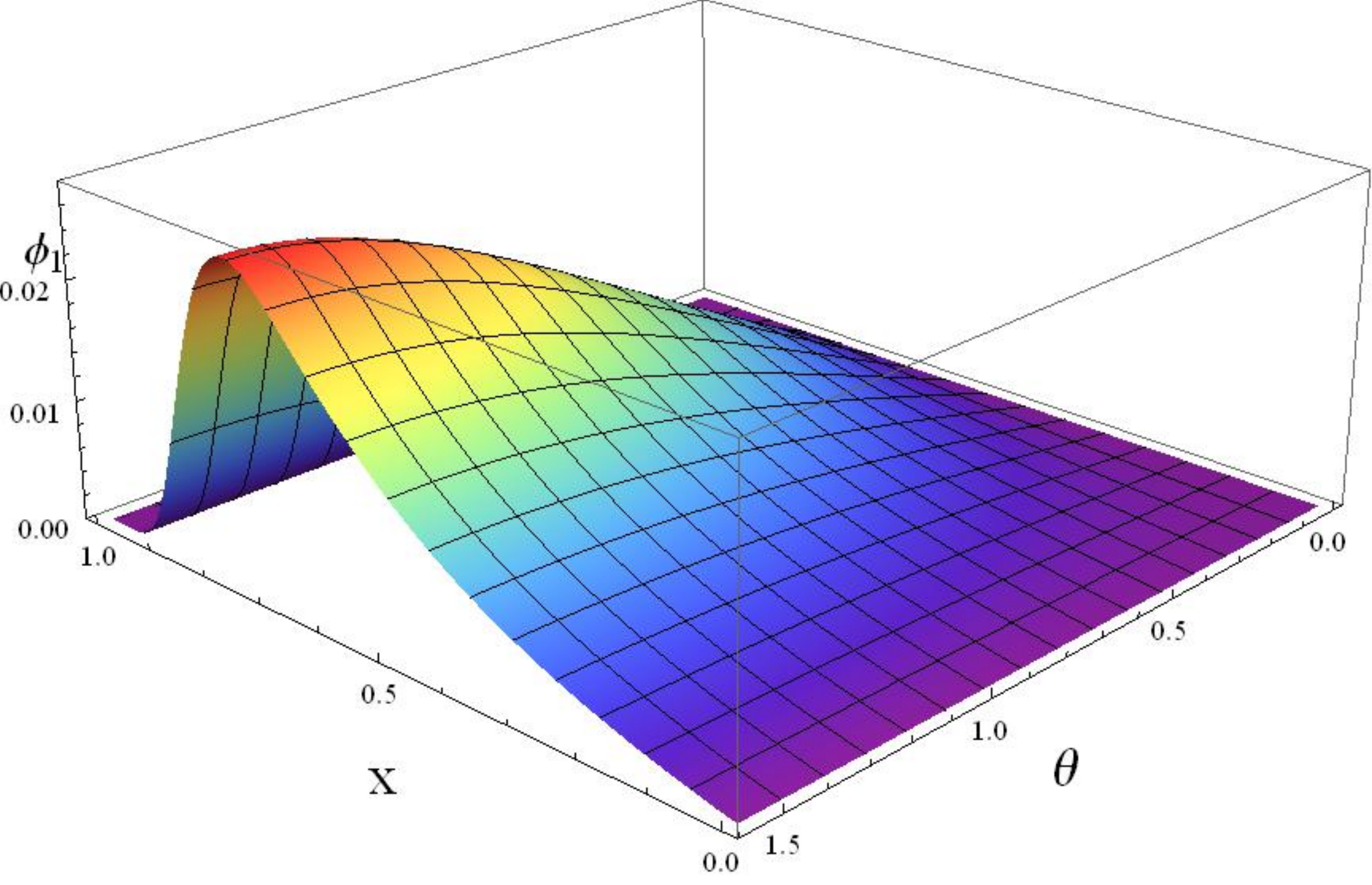}
  \end{minipage}
    \hfill
    \begin{minipage}[t]{0.3\textwidth}
    \includegraphics[width=\textwidth]{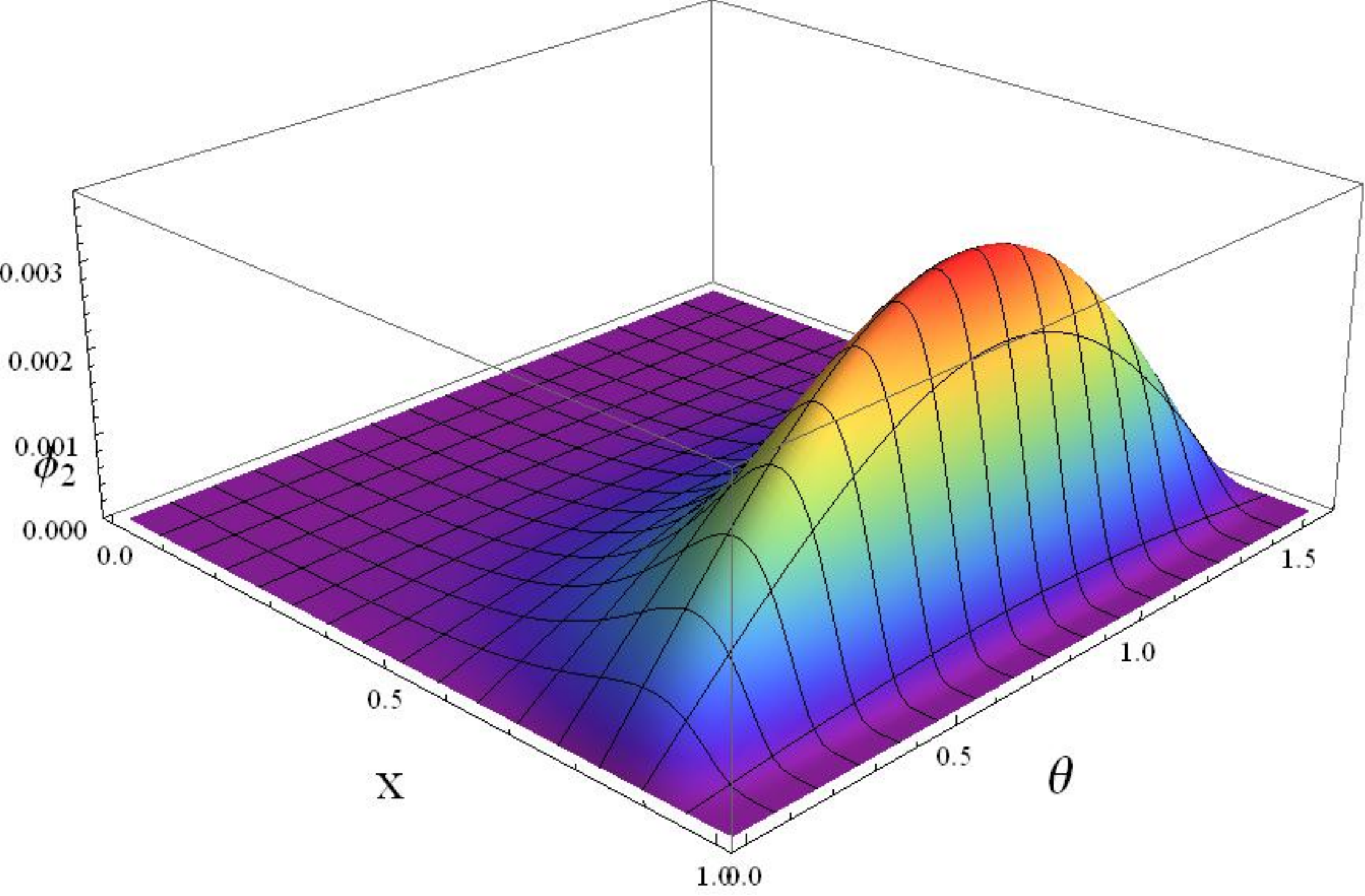}
  \end{minipage}
  \hfill
    \begin{minipage}[t]{0.32\textwidth}
    \includegraphics[width=\textwidth]{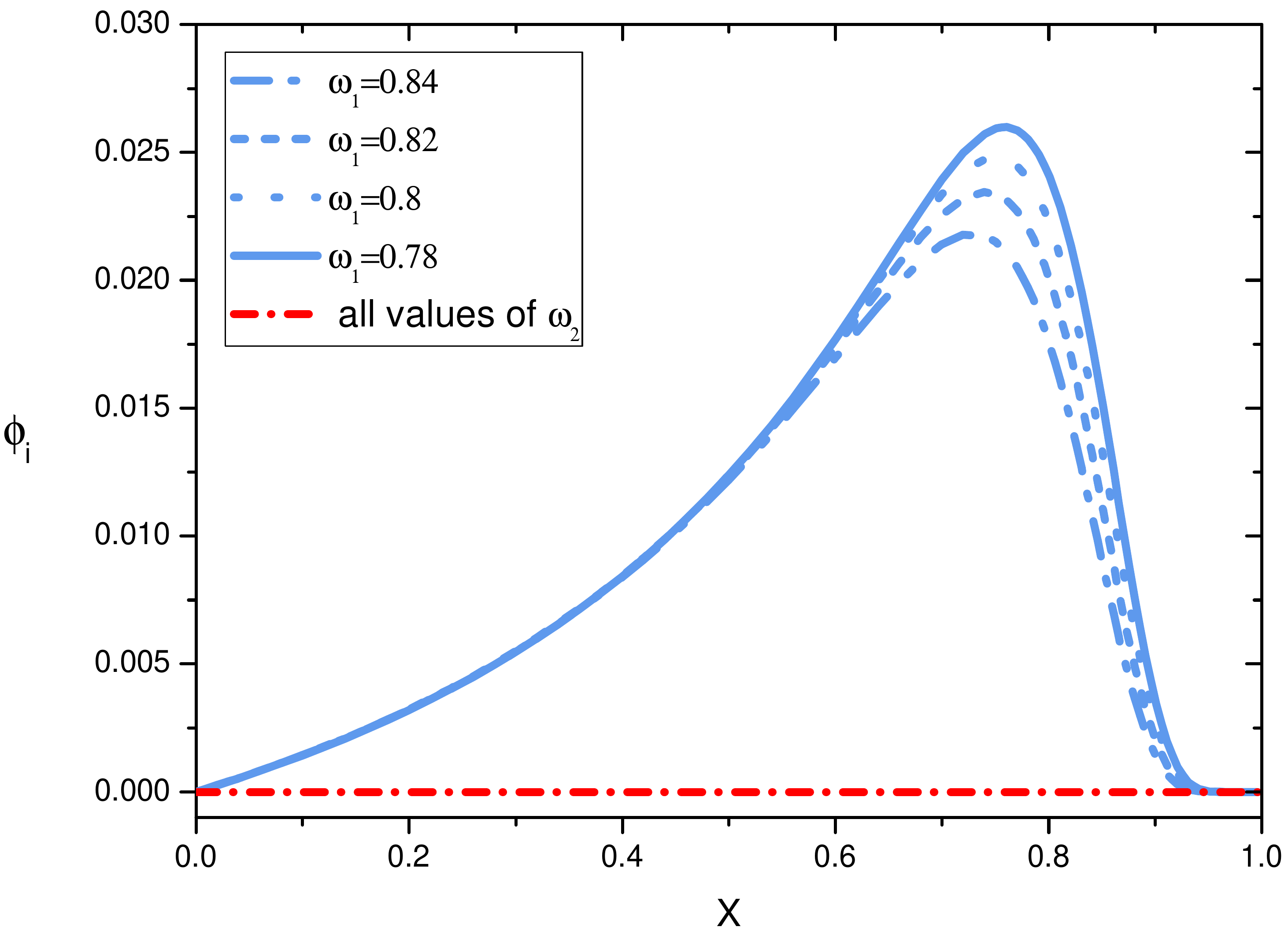}
  \end{minipage}
   \caption{ The distribution of the scalar field $\phi_1$ as a function of $x$ and $\theta$ (left panel) as well as the scalar  $\phi_2$ as a function of $x$ and $\theta$ (middle panel) with the same frequency $\omega_1=\omega_2=0.84$. The  numerical solutions of the scalar fields $\phi_i$  $(i=1,2)$  versus the boundary $x$ with an azimuthal harmonic index $m_1=1$, represented by the blue and red lines (right panel), respectively. All solutions have $m_1=m_2=1$, $\mu_1=1$, and $\mu_2=0.93$.}
  \label{fig:f-21}
\end{figure}

\begin{figure}[t]
\centering
  \begin{minipage}[t]{0.483\textwidth}
    \includegraphics[width=\textwidth]{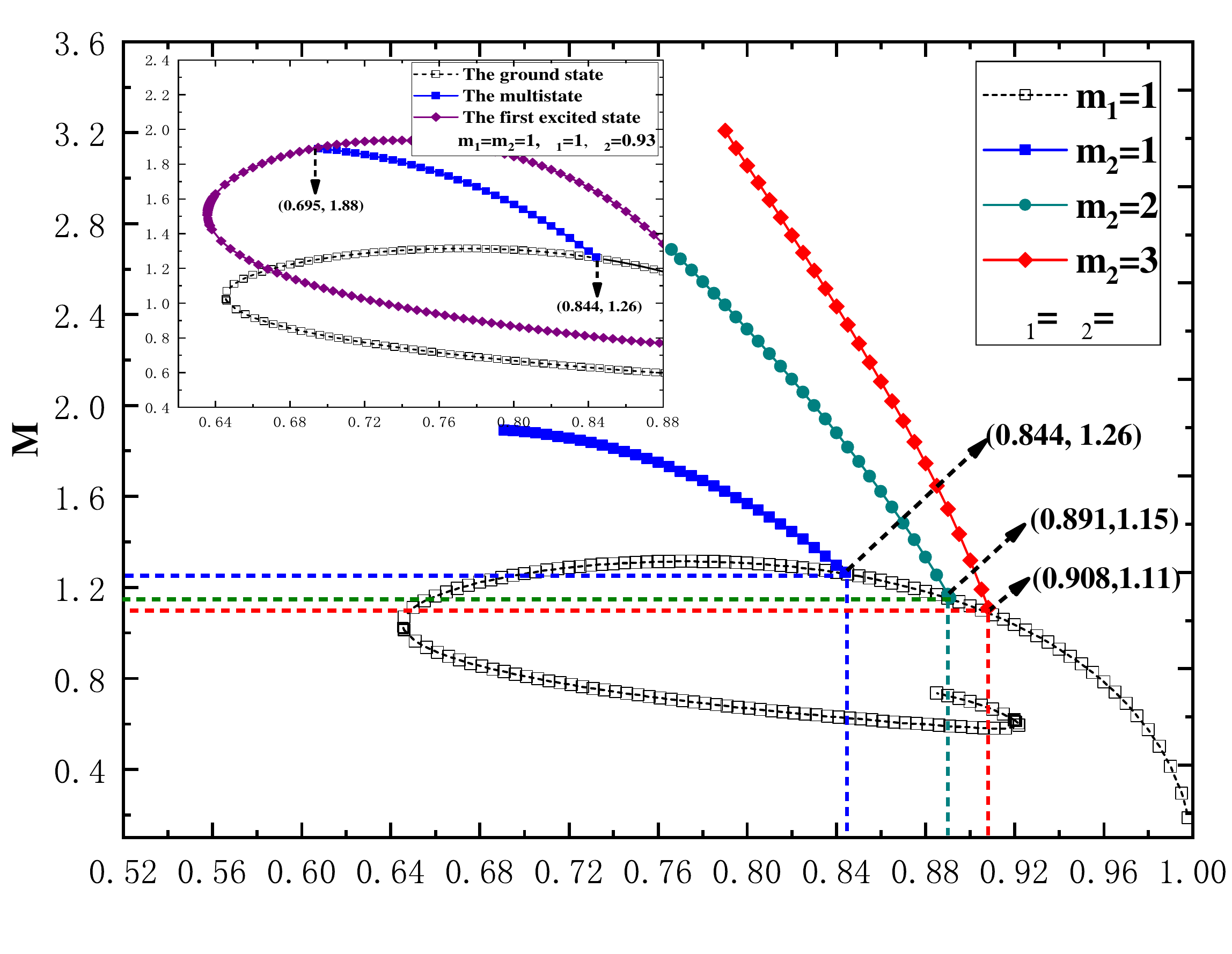}
  \end{minipage}
    \hfill
    \begin{minipage}[t]{0.48\textwidth}
    \includegraphics[width=\textwidth]{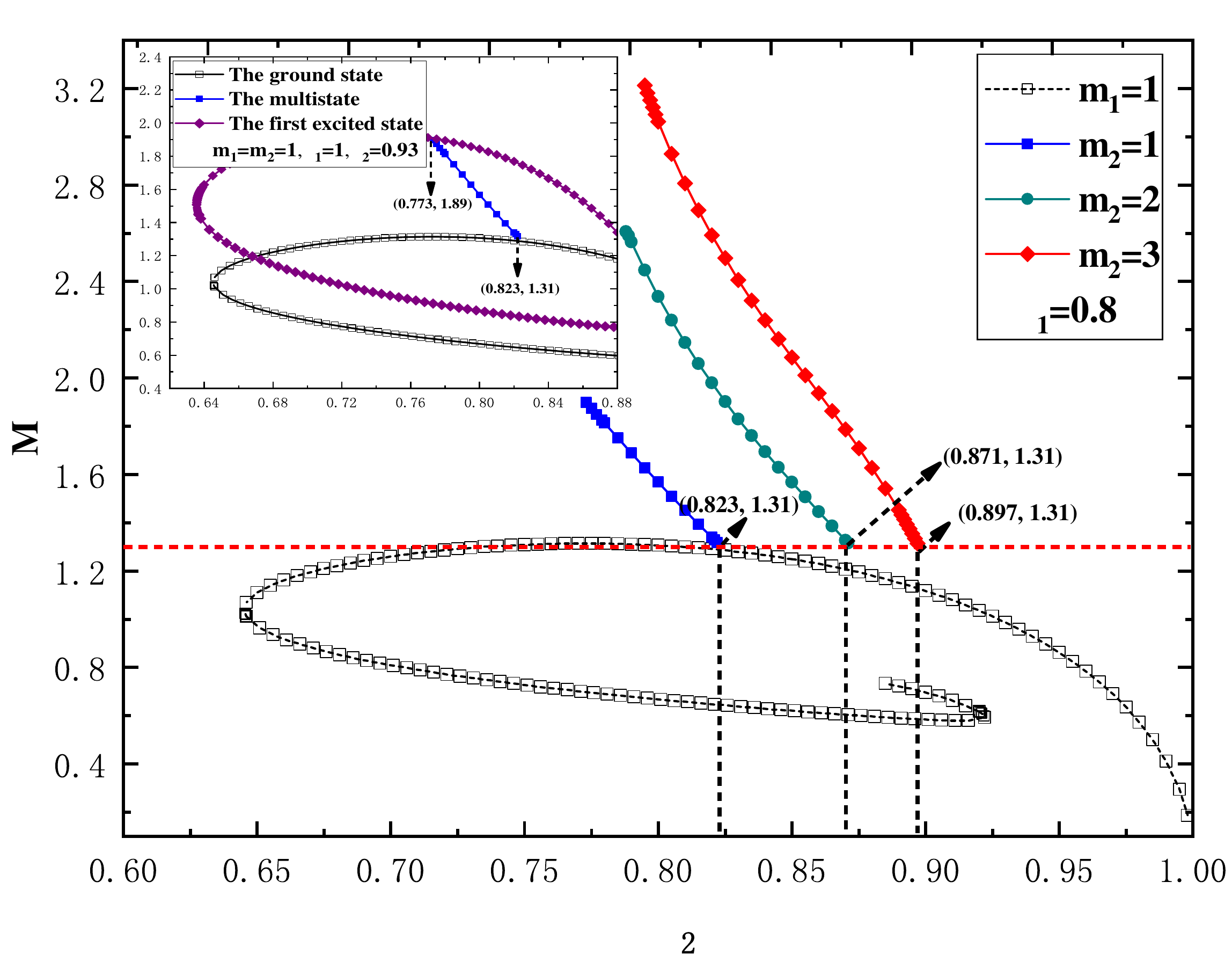}
  \end{minipage}
   \caption{\textit{Left}: The mass $M$ of  the RMSBS as a function of the synchronized frequency $\omega$ with the azimuthal harmonic indexes $m_{2}=1~,2,~3$.  Three intersection points correspond to the coordinates $(0.844,~1.26)$, $(0.891,~1.15)$, and $(0.908,~1.11)$, respectively. \textit{Right}: The mass $M$ of the RMSBS  against the nonsynchronized frequency $\omega_2$ with  the nonsynchronized frequency $\omega_1=0.8$.
   The horizon red dashed line indicate the mass $M=1.31$, and the right ends of the blue, cyan, and red dotted lines correspond to the  same value of  $M$ with coordinates $(0.823,~1.31)$, $(0.871,~1.31)$, and $(0.897,~1.31)$, respectively. In both panels, the black hollowed line indicates the ground state solutions, and the inset of  both panels show the ground state with $\mu_1=1$ (black line),  the first excited state with $\mu_2=0.93$ (purple line), and the multistate with $\mu_1=1,\mu_2=0.93$ (blue line). All solutions have $m_1=1$.}
  \label{fig:f-22}
\end{figure}

In Fig.~\ref{fig:f-21}, we exhibit the scalar fields $\phi_1$  (left panel) and $\phi_2$  (middle panel) as a function of $x$ and $\theta$ with the azimuthal harmonic indexes $m_1=m_2=1$, $\omega_1=\omega_2=0.84$, and $\mu_1=1$ and $\mu_2=0.93$.
The distribution of the scalar field $\phi_1$ (blue lines) and  $\phi_2$ (red lines)
versus the boundary $x$ for several values of the frequency $\omega_1=\omega_2$ are exhibited in the right panel of  Fig.~\ref{fig:f-21}. Along the equatorial plane at $\theta=\pi/2$,  we can see that the value of the scalar field $\phi_2$ tends to zero from the center of the boson stars to the boundary.
To discuss the properties of the RMSBS with  the $^1S^2P$ state, we mainly exhibit in  Fig.~\ref{fig:f-22}
the mass $M$  of the RMSBS versus the synchronized frequency $\omega$ and the nonsynchronized frequency $\omega_2$ with the azimuthal harmonic indexes $m_{2}=1~,2,~3$.

In the left panel of Fig.~\ref{fig:f-22}, we  show  the mass of the RMSBS versus the synchronized frequency $\omega$ with  $m_{2}=1~,2,~3$, represented by the blue, cyan, and red lines, respectively, and the black hollowed line indicates the ground state solutions for $m_1=1$.
We found that the  domain of existence of the RMSBS  are similar to the ground state boson stars in Ref. \cite{Herdeiro:2014goa}.
Again, the mass of the RMSBS  increases as the synchronized frequency $\omega$ decreases.
The RMSBS case exhibits only a stable branch with $m_{2}=1~,2,~3$,
which is similar to the  multistate with the $^1S^2S$ state.
Besides, we note that, as the azimuthal harmonic index $m_2$ increases, this maximum value of the synchronized frequency $\omega$ decreases, and the minimum value of the mass of the RMSBS  decreases as well. Hence, three sets of the RMSBS  intersect with the ground state solutions for the coordinates $(0.844,~1.26)$, $(0.891,~1.15)$, and $(0.908,~1.11)$, respectively.
From the inset in the left panel of Fig.~\ref{fig:f-22}, we note that the ground  state with $\mu_1=1$ (the black lines), and the first excited states with $\mu_2=0.93$ (the purple lines).
The multistate with $\mu_1=1,\mu_2=0.93$ (the blue line) intersect with the ground  and first excited states with  coordinates $(0.844,~1.26)$ and $(0.695,~1.88)$, respectively, which is similar to the  multistate curves with the $^1S^2S$ state.

In the right panel of Fig.~\ref{fig:f-22}, we plot the mass of the RMSBS versus the nonsynchronized frequency $\omega_2$ for the $\omega_1=0.8$.
We found that the mass of the RMSBS is increasing as $m_2$ increases.
Moreover, by increasing $\omega_2$,  the minimum value of the mass of the RMSBS is heavier than the case of the right panel of Fig.~\ref{fig:f-12}.
Furthermore, from the inset in the right panel of Fig.~\ref{fig:f-22}, we also note that the detail of the curves of the ground state with $\mu_1=1$ (black) and  the first excited states with $\mu_2=0.93$ (purple). The multistate curve with $\mu_1=1,\mu_2=0.93$ (blue) intersects with the ground  and first excited states with  coordinates $(0.773,~1.89)$ and $(0.823,~1.31)$, which is  similar to the $^1S^2S$ state case.
Because we fixed the value of $\omega_1=0.8$,  the minimal mass of the RMSBS is always a constant value $M=1.31$ for the different azimuthal harmonic indexes $m_2=2,~3$.
Thus, three coordinates correspond to $(0.823,~1.31)$, $(0.871,~1.31)$, and  $(0.897,~1.31)$, respectively.

\begin{table*}[htp]
\caption{~~The domain of existence of the  mass $\mu_2$ of the scalar field $\phi_2$ in three different situations:   The synchronized frequency $\omega=\omega_1=\omega_2$ (left panel), the nonsynchronized frequency $\omega_1$ (middle panel) and  the nonsynchronized frequency $\omega_2$ (right panel) with  $m_2=1,2,3$, respectively. In the middle and right panels, we
set $\omega_1=0.8$, $\omega_2=0.8$, respectively. All solutions have $\mu_1=1$ and $m_1=1$.}
\centering
\subtable[
]{
\scalebox{0.62}{
       \begin{tabular}{|c|c|c|c|}
       \midrule[1.0pt]\midrule[1.0pt]
\diagbox{$\omega_{~}$}{$\mu_{2}$}{$m_{2}$}&$1$&$2$&$3$\\ \midrule[1.0pt]
$0.74$&$0.837\sim0.950$&$0.771\sim0.918$&$0.757\sim0.910$\\
\hline
$0.78$&$0.881\sim0.962$&$0.819\sim0.936$&$0.811\sim0.932$\\
\hline
$0.82$&$0.914\sim0.972$&$0.864\sim0.951$&$0.844\sim0.945$\\
\hline
$0.84$&$0.928\sim0.976$&$0.885\sim0.958$&$0.864\sim0.952$\\
\hline
$0.86$&$0.940\sim0.980$&$0.904\sim0.964$&$0.885\sim0.959$\\
\midrule[1.0pt]\midrule[1.0pt]
\end{tabular}}
       \label{tab:fourthtable}
}
\subtable[
]{
\scalebox{0.615}{
       \begin{tabular}{|c|c|c|c|}
       \midrule[1.0pt]\midrule[1.0pt]
\diagbox{$\omega_{1}$}{$\mu_{2}$}{$m_{2}$}&$1$&$2$&$3$\\ \midrule[1.0pt]
$0.75$&$0.930\sim1.026$&$0.843\sim0.989$&$0.825\sim0.978$\\
\hline
$0.80$&$0.898\sim0.967$&$0.842\sim0.943$&$0.823\sim0.936$\\
\hline
$0.82$&$0.887\sim0.946$&$0.840\sim0.927$&$0.822\sim0.921$\\
\hline
$0.84$&$0.876\sim0.926$&$0.837\sim0.910$&$0.821\sim0.906$\\
\hline
$0.86$&$0.865\sim0.908$&$0.834\sim0.895$&$0.819\sim0.891$\\
\midrule[1.0pt]\midrule[1.0pt]
\end{tabular}}
       \label{tab:fifthtable}
}
\subtable[
]{
\scalebox{0.615}{
       \begin{tabular}{|c|c|c|c|}
       \midrule[1.0pt]\midrule[1.0pt]
\diagbox{$\omega_{2}$}{$\mu_{2}$}{$m_{2}$}&$1$&$2$&$3$\\ \midrule[1.0pt]
$0.78$&$0.870\sim0.941$&$0.818\sim0.919$&$0.799\sim0.912$\\
\hline
$0.80$&$0.898\sim0.967$&$0.842\sim0.943$&$0.822\sim0.936$\\
\hline
$0.82$&$0.927\sim0.994$&$0.866\sim0.968$&$0.844\sim0.961$\\
\hline
$0.84$&$0.956\sim1.021$&$0.891\sim0.993$&$0.866\sim0.985$\\
\hline
$0.86$&$0.985\sim1.048$&$0.916\sim1.018$&$0.889\sim1.009$\\
\midrule[1.0pt]\midrule[1.0pt]
\end{tabular}} \label{tab:sixthtable}
}\label{tab:table2}
\end{table*}
In  Table \ref{tab:table2}, we present the domain of existence of  the mass $\mu_2$ of the scalar field $\phi_2$ in three different situations. The mass $\mu_2$  versus the synchronized frequency $\omega$  in Table \ref{tab:fourthtable}, the nonsynchronized frequency $\omega_1$ in Table \ref{tab:fifthtable} and the nonsynchronized frequency $\omega_2$  in Table \ref{tab:sixthtable}. The three subtables have the same azimuthal harmonic index parameters $m_2=1~,2,~3$, respectively.
In Tables \ref{tab:fourthtable} and \ref{tab:sixthtable}, the domain of existence of the mass $\mu_2$ with the azimuthal harmonic index parameters $m_2=1~,2,~3$ for the same parameters $\omega=\omega_2=0.84~,0.86,~0.88$ are shown.
From Tables \ref{tab:fourthtable} and \ref{tab:sixthtable}, we obvious that the domain of existence of the mass $\mu_2$ decreases with increasing synchronized frequency $\omega$ and nonsynchronized frequency $\omega_2$.

In addition,  as  the value of the azimuthal harmonic index parameter $m_2$ increases, the domain of existence of the mass $\mu_2$ also increases.
In order to compare with the results of the domain of existence of the mass $\mu_2$ of the scalar field $\phi_2$ versus the synchronized frequency $\omega$, in Table \ref{tab:fifthtable},  we exhibit the domain of existence of the mass $\mu_2$  with the nonsynchronized frequency $\omega_1$  for the azimuthal harmonic index parameters $m_2=1~,2,~3$.
We note that, for the  values of the nonsynchronized frequencies $\omega_1=0.82$, $\omega_1=0.84$, and $\omega_1=0.86$ in Table  \ref{tab:fifthtable}, the domain of existence of the mass $\mu_2$ has the similar behavior as the case of $^1S^2S$ state in Table \ref{tab:secondtable}.

\section{Conclusion}\label{sec5}

In this paper, we  have constructed and analyzed rotating boson stars composed of the  coexisting states of two massive scalar fields,
including the ground state and the first excited state. Comparing with the solutions of the rotating ground state boson stars in Ref. \cite{Herdeiro:2014goa},
we have found that  the RMSBS have two types of nodes, including the $^1S^2S$  state and  the $^1S^2P$  state.
By calculating the coexisting phase of the RMSBS for the two types of nodes,  we found that the domain of existence of the mass $\mu_2$ decreases with an increasing synchronized frequency $\omega$, meanwhile,  by increasing the value of the azimuthal harmonic index parameter $m_2$, the mass domain as the synchronized frequency $\omega$ keeps increasing.
Furthermore, when the  nonsynchronized frequency $\omega_2$ increases, the scalar field  of the first excited state could decease to zero, and the minimal mass of the RMSBS is provided by the scalar field of the ground state. Therefore, the mass of the RMSBS is always a constant value for the different azimuthal harmonic indexes $m_2=1~,2,~3$. In addition, from the numerical results, it is obvious that the mass of the RMSBS is heavier than the case of the ground state.

In order to better understand  the stability properties of the RMSBS, according to the numerical analysis of  the stability of the excited boson stars studied in \cite{Collodel:2017biu}, the authors found the most stable solution will always belong to the set of ground state solutions, and  for the case of nonrotating  multistate boson stars \cite{Bernal:2009zy}, the authors also found that there is a region of the solution space with stable configurations, that is, the deeper gravitational potential generated by the ground state, which is large enough to stabilize the excited state. It is worth to point out that it is difficult to numerically analyze the stability properties of rotating multistate boson stars. However, a good way to guarantee the stability of a specific solution is to have the linear perturbation mode in \cite{Ganchev:2017uuo}.

Recently, motivation by the increasing interest in the models which consider scalar fields as viable dark matter candidates
\cite{Sahni:1999qe,Matos:2000ng,Hu:2000ke,Bernal:2016lll,Padilla:2019fju} has increased. For the ground state boson stars, these structures could produce rotation curves (RC), but the RC are not flat enough at large radii. Moreover, the excited state boson stars typically produce a more physically realistic, flatter RC, for which the solutions are unstable. In 2010, Bernal et al. \cite{Bernal:2009zy} have obtained the configurations with two states, a ground and a first existed state, and they have demonstrated
that the RC of multistate boson stars are flatter at large radii than the RC of single boson stars. As discussed above, the case of multistate boson stars is the spherically symmetric, nonrotating solutions. For the axisymmetric, rotating multistate solutions, however, we believe that the RC of rotating multistate boson stars also are flatter at large radii than the RC of single rotating solutions, and we hope that the RC of rotating multistate boson stars could be better used to  accurately fit the rotation curves within the observational data. Therefore, we will calculate the rotation curves of rotating multistate boson stars in future  work.

There are several interesting extensions of our work. Firstly, we have studied the rotating multistate boson stars; we would like to investigate how self-interactions of the scalar field affects the rotating multistate boson stars inspired by the work \cite{Kleihaus:2011sx}.
Secondly, the extension of our study is to construct the multistated Kerr black hole with scalar hairs, where two coexisting states of the scalar field are presented, including the ground and excited states. Finally, we are planning to numerically analyze of the linear stability properties of the rotating multistate boson stars in  future work.

\section*{Acknowledgement}
YQW would like to thank  Yu-Xiao Liu  and Jie Yang for  helpful discussions. We would also like to thank the anonymous
referee for the valuable comments which helped to improve the manuscript.  Some computations were performed on the   Shared Memory system at  Institute of Computational Physics and Complex Systems in Lanzhou University. This work was supported by the  Natural Science Foundation of China (Grants No. 11675064,  No. 11522541, and No. 11875175), and the Fundamental Research Funds for the Central Universities (Grants No. lzujbky-2017-182, No. lzujbky2017-it69, and  No. lzujbky-2018-k11).

\providecommand{\href}[2]{#2}\begingroup\raggedright
\endgroup
\end{document}